\documentclass[conference]{IEEEtran}
\IEEEoverridecommandlockouts
\usepackage{cite}
\usepackage{amsmath,amssymb,amsfonts}
\usepackage{algorithmic}
\usepackage{graphicx}
\usepackage{textcomp}
\usepackage{xcolor}
\usepackage{listings}

\usepackage{comment}
\usepackage{listings}
\usepackage{color}
\definecolor{codegreen}{rgb}{0,0.6,0}
\definecolor{codegray}{rgb}{0.5,0.5,0.5}
\definecolor{codepurple}{rgb}{0.58,0,0.82}
\definecolor{backcolour}{rgb}{0.95,0.95,0.92}
\definecolor{azure}{rgb}{0.0, 0.5, 1.0}
\definecolor{blue(ryb)}{rgb}{0.01, 0.28, 1.0}
\definecolor{blue(pigment)}{rgb}{0.2, 0.2, 0.6}
\usepackage[T1]{fontenc}
\usepackage{tabu,multirow}

\def\BibTeX{{\rm B\kern-.05em{\sc i\kern-.025em b}\kern-.08em
    T\kern-.1667em\lower.7ex\hbox{E}\kern-.125emX}}
\begin{document}

\title{A Mixed-Precision RISC-V Processor for Extreme-Edge DNN Inference\\
\vspace{10pt}
\thanks{This  work  was  supported  in  part  by  OPRECOMP  (Open  trans-PREcisionCOMPuting) Grant Agreement No. 732631, and WiPLASH ( Wireless Plas-ticity for Heterogeneous Massive Computer Architectures) Grant AgreementNo. 863337. Both projects are funded from the European Union’s Horizon 2020 research and innovation program.}
}

\author{\IEEEauthorblockN{Gianmarco Ottavi\IEEEauthorrefmark{2}, Angelo Garofalo\IEEEauthorrefmark{2}, Giuseppe Tagliavini \IEEEauthorrefmark{2}, Francesco Conti\IEEEauthorrefmark{2}\IEEEauthorrefmark{1}, Luca Benini\IEEEauthorrefmark{2}\IEEEauthorrefmark{1} and Davide Rossi\IEEEauthorrefmark{2}} 
\IEEEauthorblockA{\textit{DEI, University of Bologna, Italy}\IEEEauthorrefmark{2} \quad \textit{IIS lab, ETH Zurich, Switzerland}\IEEEauthorrefmark{1} \\
$\{$gianmarco.ottavi2, davide.rossi, angelo.garofalo, giuseppe.tagliavini$\}$@unibo.it\\ $\{$fconti, lbenini$\}$@iis.ee.ethz.ch}}
 
\maketitle

\begin{abstract}
Low bit-width Quantized Neural Networks (QNNs) enable deployment of complex machine learning models on constrained devices such as microcontrollers (MCUs) by reducing their memory footprint.
Fine-grained asymmetric quantization (i.e., different bit-widths assigned to weights and activations on a tensor-by-tensor basis) is a particularly interesting scheme to maximize accuracy under a tight memory constraint~\cite{rusci2019memory}.
However, the lack of sub-byte instruction set architecture (ISA) support in SoA microprocessors makes it hard to fully exploit this extreme quantization paradigm in embedded MCUs.
Support for sub-byte and asymmetric QNNs would require many precision formats and an exorbitant amount of opcode space.
%
In this work, we attack this problem with status-based SIMD instructions: rather than encoding precision explicitly, each operand's precision is set dynamically in a core status register. We propose a novel RISC-V ISA core \textit{MPIC} (Mixed Precision Inference Core) based on the open-source RI5CY core. Our approach enables full support for mixed-precision QNN inference with different combinations of operands at \mbox{16-}, \mbox{8-}, \mbox{4-} and 2-bit precision, without adding any extra opcode or increasing the complexity of the decode stage. 
Our results show that MPIC improves both performance and energy efficiency by a factor of 1.1--4.9$\times$ when compared to software-based mixed-precision on RI5CY; with respect to commercially available Cortex-M4 and M7 microcontrollers, it delivers 3.6--11.7$\times$ better performance and 41--155$\times$ higher efficiency.
\end{abstract}

\section{Introduction}
\label{sec:introduction}
Running complex applications on embedded systems like microcontrollers (MCUs) requires optimization on both software and hardware due to severe constraints in terms of memory size, power consumption, and computing power. In an Internet-of-Things (IoT) environment, wireless communication to higher-level nodes often dominates the power budget. Algorithms such as Deep Neural Networks (DNNs), more specifically Convolutional Neural Networks (CNNs) which are state-of-the-art for computer vision and speech recognition, are used in computing at the edge of IoT to reduce the amount of data to transmit by communicating only classes or high-level features instead of the raw sensor data. The complexity of these algorithms typically requires millions of Multiply-Accumulate (MAC) operations and significant memory footprint, where memory is a valuable resource due to its cost in terms of area and power.
%

An effective way to reduce the memory footprint of DNNs is \textit{quantization}, a technique that reduces inputs and weights to fixed-point formats such as 8- bits, and even sub-byte like 4- and 2- bits \cite{rusci2019memory,moons2017minimum,hubara2017quantized}.
Banner \textit{et al.} proposed a methodology to quantize both weights and activations to 4-bit with an accuracy drop of only a few percent, not modifying the training and not requiring a full dataset. Rusci \textit{et al.} \cite{rusci2019memory} show how, using mixed-precision quantization for each layer, it is possible to reduce by up to 7$\times$ the memory footprint of DNNs, incurring only in a 4\% accuracy loss. However, although quantization provides a clear reduction of memory bandwidth visible also in general-purpose processors \cite{Clover}, much of the inference-time benefit is accessible only through customized hardware accelerators \cite{Marian} or with an FPGA implementation of quantized arithmetic units \cite{Deeper}. To the best of the authors' knowledge, the only recent work taking advantage of quantized formats in software processors is the one presented by Anderson \textit{et al.}\cite{SAMD}. It proposed a software technique exploiting arbitrary bit-precise signed and unsigned integer operations embedding a vector architecture with custom bit-width lanes in fixed-width scalar arithmetic \cite{SAMD}. However, this comes with significant effort in application porting.

From the hardware perspective, the only relevant research work in this field is the reconfigurable Parallel Balanced-Bit-Serial (PBBS) vector processing tile presented by Wu \textit{et al.} \cite{BitSerial}, which is suitable for improving the efficiency of sub-byte single instruction multiple data (SIMD) computations of heavily leakage-dominated ULP designs. However, code serialization significantly degrades performance and efficiency in near- and super-threshold operating points. On the other hand, the totality of commercial MCUs operates at the finest granularity of 1-byte data \cite{ARMV8-ISA,riscv-spec}. The new ARM \cite{M55} ISA specialized for machine learning, implemented by the \textit{Cortex M55} processor, enhances the ARMv8 with extensions similar to the ones presented in \cite{RI5CY}, such as 8-bit SIMD instructions, loops, and conditional execution extensions. In addition, it provides pipelined execution of load and mac instructions \cite{M55} that allows maximizing utilization of MAC units during the execution of regular patterns (e.g., convolutions). 

However, similarly to all other commercial cores, the ARM \textit{Cortex M55} does not support natively smaller than 8-bit SIMD instructions. Hence, data have to be presented as a byte for computation, even if it is ``packed'' in a more compact representation. First, this means that there is no way to exploit the additional parallelism because the datapath is hardwired to 8 bits. Second, in the tight inner loops of the quantized DNN kernels, the cost of unpacking and packing data can be extremely high, leading to up to $2.5\times$ worse performance than directly using 8-bit data, as shown in the results. In our experiments, sub-byte and mixed-precision quantization by itself improved only the implementation feasibility of a network in MCUs (in term of squeezing the network memory footprint), but not its performance and efficiency in computation \cite{garofalo2020pulp}.

Supporting many different precision formats to avoid data unpacking can be challenging in a general-purpose MCU, because it leads to the proliferation in the number of instructions, saturating the encoding space. Variable-length instructions offer a potential solution to this problem, but only at the cost of code bloat and increased complexity of the decoding stage, which would result in a significant penalty for what concerns the power consumed by the MCU \cite{azizi2010energy}. In this work, we propose a lightweight processor specialization for quantized DNNs leveraging a status-based approach to counter the proliferation of SIMD instructions necessary to support mixed-precision computations. Such instructions do not encode precision explicitly; rather, they encode a ``virtual'' SIMD instructions, which contain no precision information and the latter is specialized at run-time by setting the precision of the operands in a core status register. In this way, the same virtual instructions can encode a range of operand precision, enabling much higher code efficiency.

The main contributions of this paper are the following:
first, we introduce \textit{XMPI}, a RISC-V ISA extension introducing mixed-precision and heavily quantized SIMD instructions to boost the execution of Quantized Neural Network workloads from 16 down to 2 bits;
moreover, we extend the functionality of the RI5CY core \cite{RI5CY}, a state-of-the-art open-source RISC-V core, to support status-based operation. We call this new core \textit{MPIC} (Mixed Precision Inference Core). We then integrated XMPI and added new execution stage functional units to operate at the granularity of 2 and 4 bits.

To validate our design, we deployed the new core into PULPissimo~\cite{schiavone2018quentin}, a single-core open-source MCU of the PULP\footnote{https://pulp-platform.org/} family~\cite{PULP}. We implemented the full layout on a commercially available technology at 22nm FDX from Globalfounderies to evaluate overheads in terms of power, area, and frequency with respect to the baseline RI5CY core. We benchmarked the mixed-precision extended core against RI5CY, ARM Cortex M7, and ARM Cortex M4 cores on a QNN layer with different quantization configurations. The new approach of MPIC avoided the encoding of 200 new instructions keeping power consumption on the level of the baseline RI5CY core. Our results show that the new ISA brings 1.1--4.9$\times$ better performance and energy efficiency when compared to software-based mixed-precision on the RI5CY core; moreover, we also compare with commercially available MCUs based on Cortex-M7 and M4 cores, showing that our solution provides a boost of 3.6--11.7$\times$ in performance and 41--155$\times$ in energy efficiency.
\section{Background}
\subsection{RI5CY Core}
RI5CY, used as a baseline for the proposed work, is an open-source core featuring a 4 stage in-order single-issue pipeline based on the RISC-V ISA~\cite{riscv-spec}. It supports the standard RISC-V extensions (I, M, C, and F) but also includes a non-standard extension, called \textit{XpulpV2}, that introduces several features such as hardware loops, bit manipulation instructions, load/store post-increment instructions, SIMD for 16- and 8-bit format (more information can be found at~\cite{RI5CY}). As later described in Section~\ref{bench}, these features provide 4.4$\times$ performance in 8-bit kernels when compared to SoA cores such as ARM Cortex M7.

\subsection{Quantized Neural Networks}
The QNN layers used for the experimental assessment of our approach adopt layer-wise linear quantization. The quantization process takes care of mapping each tensor to integers values. We have 3 categories of tensors to quantize: input feature maps, output feature maps, and weights (\textbf{x}, \textbf{y}, and \textbf{w}, respectively). A generic real-valued tensor \textbf{t} that belongs inside the range of $[\alpha_\mathbf{t}, \beta_\mathbf{t})$ can be expressed as:
\begin{equation}
\mathbf{t} = \alpha_\mathbf{t} + \varepsilon_\mathbf{t} \cdot \textsc{int}(\mathbf{t})
\label{eq:real_value_tensor}
\end{equation}
where $\textsc{int}(\mathbf{t})$ is the value of \textbf{t} mapped to an $N$-bit integer, $\varepsilon_\mathbf{t} = (\beta_\mathbf{t} - \alpha_\mathbf{t})/2^{N}$ and $\alpha_\mathbf{t}$ is the bias to shift the value back to its original range. Imposing $\alpha_\mathbf{t}=0$ for both input and output feature maps (but not weights) gives a QNN that can be trained efficiently by means of linear quantization-aware training~\cite{choi2018pact}. It is possible to work directly on quantized integer values and apply convolution, normalization and activation:
\begin{equation}
\textsc{int}(\mathbf{y}) = \mathrm{quant}\Big(\mathrm{conv}\big(\textsc{int}(\mathbf{w}), \textsc{int}(\mathbf{x})\big)\Big)
\label{eq:res_1}
\end{equation}
The result of the convolution $\phi = \mathrm{conv}\big(\textsc{int}(\mathbf{w}),\textsc{int}(\mathbf{x})\big)$ is still an integer tensor but has to be represented with a larger number of bits than its input ($\varepsilon_\phi$ is smaller than both $\varepsilon_\mathbf{x}$ and $\varepsilon_\mathbf{w}$). We then have the function $\mathrm{quant}(\cdot)$ that first applies batch-normalization (if any) to $\phi$ and then scales the result to the proper number of output bits.

Mixed-precision QNNs do not impose the same number of bits for activations and weights, opening the possibility to have more sensitive layers and/or tensors to be represented at higher precision while strongly quantizing the rest~\cite{rusci2019memory}.

\subsection{QNN Execution Model}
\label{sec:execution_model}
The execution model for QNNs adopted in this work is based on the PULP-NN  library~\cite{garofalo2020pulp}, a library composed of QNN kernels optimized to run on PULP systems. These libraries are inspired by the ARM CMSIS-NN~\cite{lai2018cmsis}, but they include additional support for sub-byte quantization of INT-4, -2, -1 integer types. For efficient execution on PULP cores, convolutional layers inference is split into three phases. The \textit{im2col} phase takes the 3D input features and maps it into a 1D vector. The \textit{MatMul} organizes the innermost dot products of the convolution operator 
as a set of 4$\times$2 matrix multiplications: each inner-loop of the convolution outputs 2 spatially adjacent pixels along 4 consecutive channels. It does so by loading two input buffers and 4 adjacent filters and leveraging the reuse of input elements to ensure a more efficient ratio of MAC/load~\cite{garofalo2020pulp}. Since the results of the matrix multiplication have a higher dimension than their inputs, the last phase (\textit{QntPack}) discretizes the 32-bit outputs of the \textit{MatMul} to their target precision and pack them into 32-bit variables. For this purpose, different discretization techniques are employed depending on the case: 8-bit output uses scaling and clamping~\cite{lai2018cmsis}, while 4- and 2-bit configurations use thresholds~\cite{hubara2017quantized,rusci2019memory}. This operation compares the result of the matrix multiplication with a set of thresholds computed at training time, which directly implement the $\mathrm{quant}$ function of Eq.~\ref{eq:res_1}.


\section{ISA EXTENSION}

\subsection{Computational Model}
The \textit{XpulpV2} extension of the RI5CY core ISA supports 16- and 8-bit SIMD operations.
Supporting formats from 16- down to 2-bit, and all the permutations of mixed-precision operations, would require 10 different encodings per each supported MAC-SIMD instruction for a total of 292 instructions versus 92 of the baseline core. This would require 4 bits, while only one is available for this purpose in the current RISC-V encoding~\footnote{https://www.pulp-platform.org/docs/ri5cy\_user\_manual.pdf}.
\begin{figure}[t]
\centerline{\includegraphics[width=0.5\textwidth]{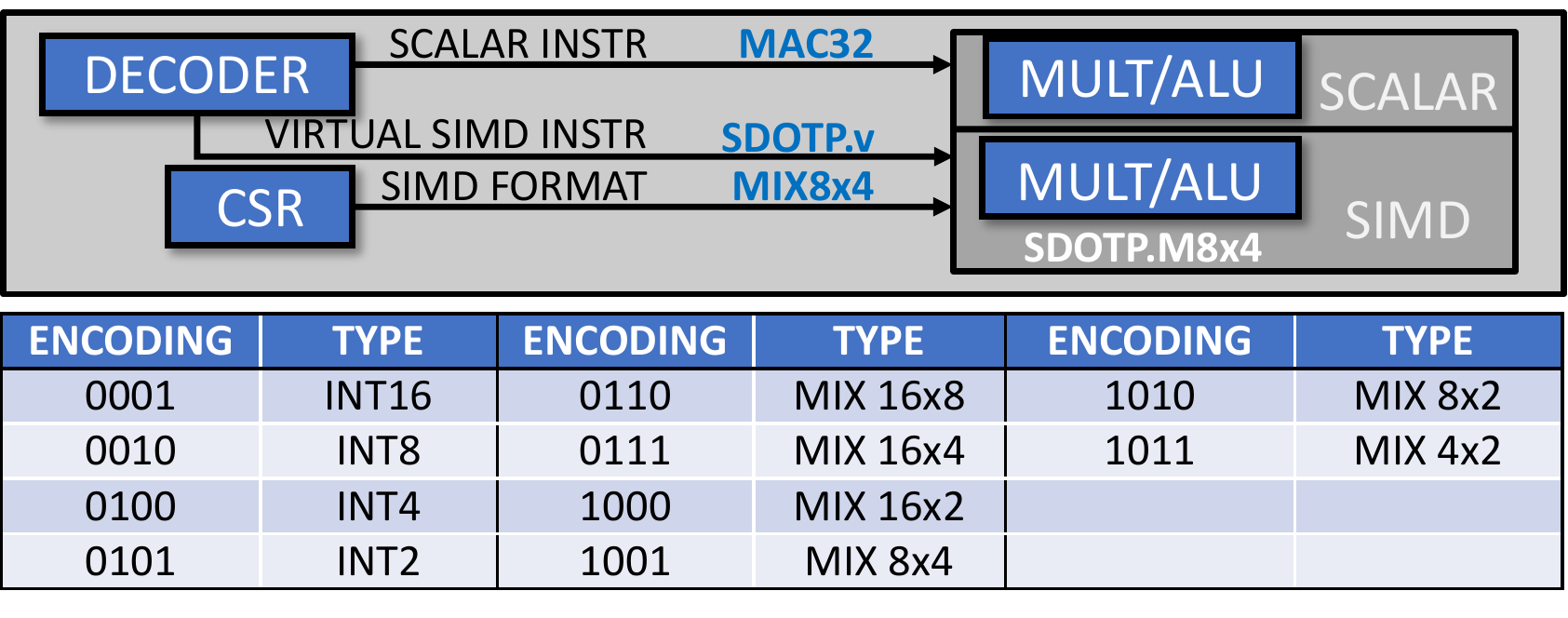}}
\caption{Control signals for SIMD and Scalar instructions. The SIMD instruction is a Sum-of-Dot-Product and the format is a mixed-precision 8x4. The bottom picture contains the encoding of the formats that are contained inside the CSR. }
\label{fig:sb-dec}
\end{figure}
In the MPIC core, we eliminate the problem of encoding space by using virtual instructions.
As depicted in Figure~\ref{fig:sb-dec}, for regular scalar instructions the decoder directly produces control signals towards the \textit{ex\_stage} (e.g., in the case of a MAC with 32-bit operands); virtual SIMD instructions require additional information from the status registers (CSR) to be specialized (e.g., in the case of an 8-bit by 4-bit sum-of-product).

\begin{figure}
    \centering
    \includegraphics[width=0.5\textwidth]{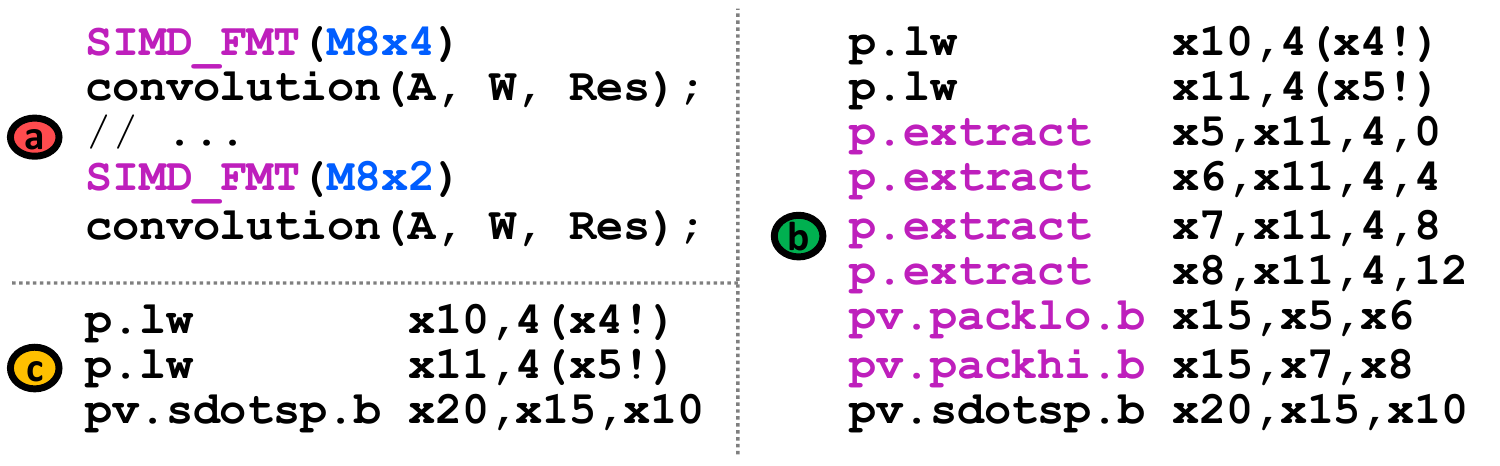}
    \caption{\textbf{a}) MPIC Functions Call changing precision before executing operations; \textbf{b}) RI5CY inner loop with data packing/unpacking overhead; \textbf{c}) MPIC inner loop with MAC instructions executed directly.}
    \label{fig:mpic_coding}
\end{figure}

Application code requires explicit modifications to use SIMD virtual instructions.
In Figure~\ref{fig:mpic_coding} \textbf{a)}, we have an example of a QNN with multiple layers using different precisions. Here we can note how before the function calls, we set the precision with the \textit{SIMD\_FMT} macro, which writes the appropriate format encodings into the CSR. If we ``zoom'' inside the functions and get to the inner loop of the \textit{8x4} mixed-precision kernel, we can see how supporting directly in hardware this new format benefits the computation. In Figure~\ref{fig:mpic_coding} \textbf{b)}, we show the normal instruction flow using RI5CY: first, we load the activations and weights from memory; then, we unpack four of the eight operands in the 32-bit register containing the current weight. Once unpacked, they have to be packed again into 8-bit operands, to take advantage of RI5CY's 8-bit vector MAC instruction. On the other hand, MPIC only requires to load and execute the vectorial MAC (Figure~\ref{fig:mpic_coding} \textbf{c)}).
Thus, it saves two-thirds of the instructions on the inner loop when running on data smaller than a byte (or mixed-precision).

\begin{figure*}
\center
  \includegraphics[width=0.95\textwidth]{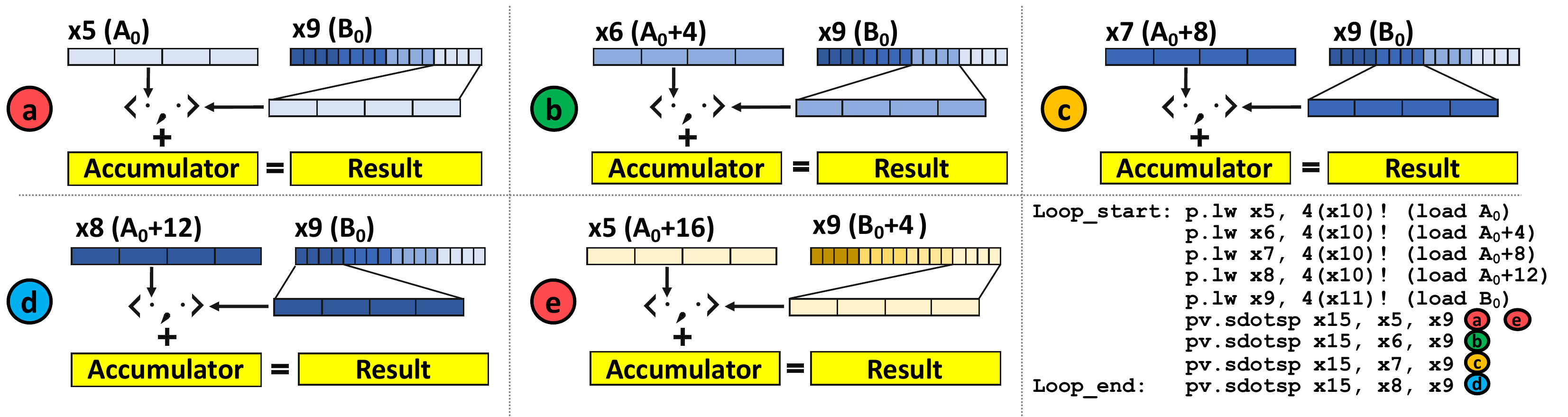}
  \caption{Matrix multiplication between operands of size 8- and 2-bit. Vector B contains four times the operands of Vector A requiring the fetch of 3 more vectors to "exhaust" Vector B. In each step, we have a group of operands unpacked (via Hardware) from Vector B, extended to match the size of Vector A, and finally execute the dot-product between the vectors where the partial result is added to the accumulator to get the final result. On the bottom right, we can see the kernel assembly: \texttt{p.lw} are post increment loads that increment by 4 the pointer after load, and \texttt{pv.sdotsp} is a signed sum of dot-product. 
  }
  \label{fig:mixed-matmul}
\end{figure*}

In Figure~\ref{fig:mixed-matmul}, we illustrate how a matrix multiplication kernel works in the case of mixed-precision operands. Having the same instructions that deal with both mixed-precision and uniform-precision operations requires added logic for the management of the input with smaller operands; by construction, we always map it to input B. As shown in Figure~\ref{fig:mixed-matmul}, input B can remain stationary for multiple MAC instructions before new data is needed to be fetched. In the 8x2-bit example of the figure, choosing the correct group of 4 operands out of the 16 is crucial for the correctness of the result. To this end, we designed a controller to deal with this problem, which is explained in more detail in Section \ref{microark}.

\subsection{Virtual Instructions}
Table~\ref{ISA_EXT_QNN} lists the instructions in the XMPI extension. The instructions are derived from a subset of \textit{XpulpV2}, where they are available in 16-, and 8-bit precision only. We extended them with additional support for symmetric 4- and 2-bit precision and, for the dot-product instructions, with also mixed-precision support. We have different versions of these instructions: the \textit{sc} variant is an operation between a scalar and a vector; the \textit{i} variant uses the value from the immediate field instead of a register; finally, we support signed and unsigned variants (dotp instructions also have a hybrid unsigned-signed).

\begin{table}[t]
\begin{center}
\scalebox{1}[1]{\begin{tabular}{|l|l|}
    \hline
    \textbf{Instruction} &  \textbf{Description}\\
    \hline
    \multicolumn{2}{|c|}{\textbf{ALU SIMD Instr.}} \\
    \hline
    pv.add[.sc(i)] &   rD[i] = rs1[i] + rs2[i] \\
    pv.sub[.sc(i)] &   rD[i] = rs1[i] - rs2[i] \\
    pv.avg(u)[.sc(i)]  &   rD[i] = (rs1[i] + rs2[i]) >> 1    \\
    \hline
    \multicolumn{2}{|c|}{\textbf{Vector Comparison Instr.}} \\
    \hline
    pv.max(u)[.sc(i)]  &   rD[i] = rs1[i] > rs2[i] ? rs1[i] : rs2[i]   \\
    pv.min(u)[.sc(i)]  &   rD[i] = rs1[i] < rs2[i] ? rs1[i] : rs2[i]   \\
    \hline
    \multicolumn{2}{|c|}{\textbf{Vector Shift Instr.}} \\
    \hline
    pv.srl[.sc(i)] &   rD[i] = rs1[i] >> rs2[i] Shift is logical   \\
    pv.sra[.sc(i)] &   rD[i] = rs1[i] >> rs2[i] Shift is arithmetic    \\
    pv.sll[.sc(i)] &   rD[i] = rs1[i] << rs2[i]    \\
    \hline
    \multicolumn{2}{|c|}{\textbf{Vector ABS Instr.}} \\
    \hline
    pv.abs  &   rD[i] = rs1[i] < 0 ? -rs1[i] : rs1[i]   \\
    \hline
    \multicolumn{2}{|c|}{\textbf{Dot Product Instr.}} \\
    \hline
    pv.dotup[.sc(i)]   &   rD = rs1[0]*rs2[0] + ... + rs1[7]*rs2[7]    \\
    pv.dotusp[.sc(i)]  &   rD = rs1[0]*rs2[0] + ... + rs1[7]*rs2[7]    \\
    pv.dotsp[.sc(i)]   &   rD = rs1[0]*rs2[0] + ... + rs1[7]*rs2[7]    \\
    pv.sdotup[.sc(i)]  &   rD = rs1[0]*rs2[0] + ... + rs1[7]*rs2[7] + rs3  \\ 
    pv.sdotusp[.sc(i)] &   rD = rs1[0]*rs2[0] + ... + rs1[7]*rs2[7] + rs3  \\
    pv.sdotsp[.sc(i)]  &   rD = rs1[0]*rs2[0] + ... + rs1[7]*rs2[7] + rs3 \\
    \hline
\end{tabular}}
\vspace{+2mm}
\caption{List of instructions extended by XMPI}
\label{ISA_EXT_QNN}
\end{center}
\vspace{-5mm}
\end{table}

\subsection{Microarchitecture}
\label{microark}
\begin{figure}[t]
\centerline{\includegraphics[width=0.5\textwidth]{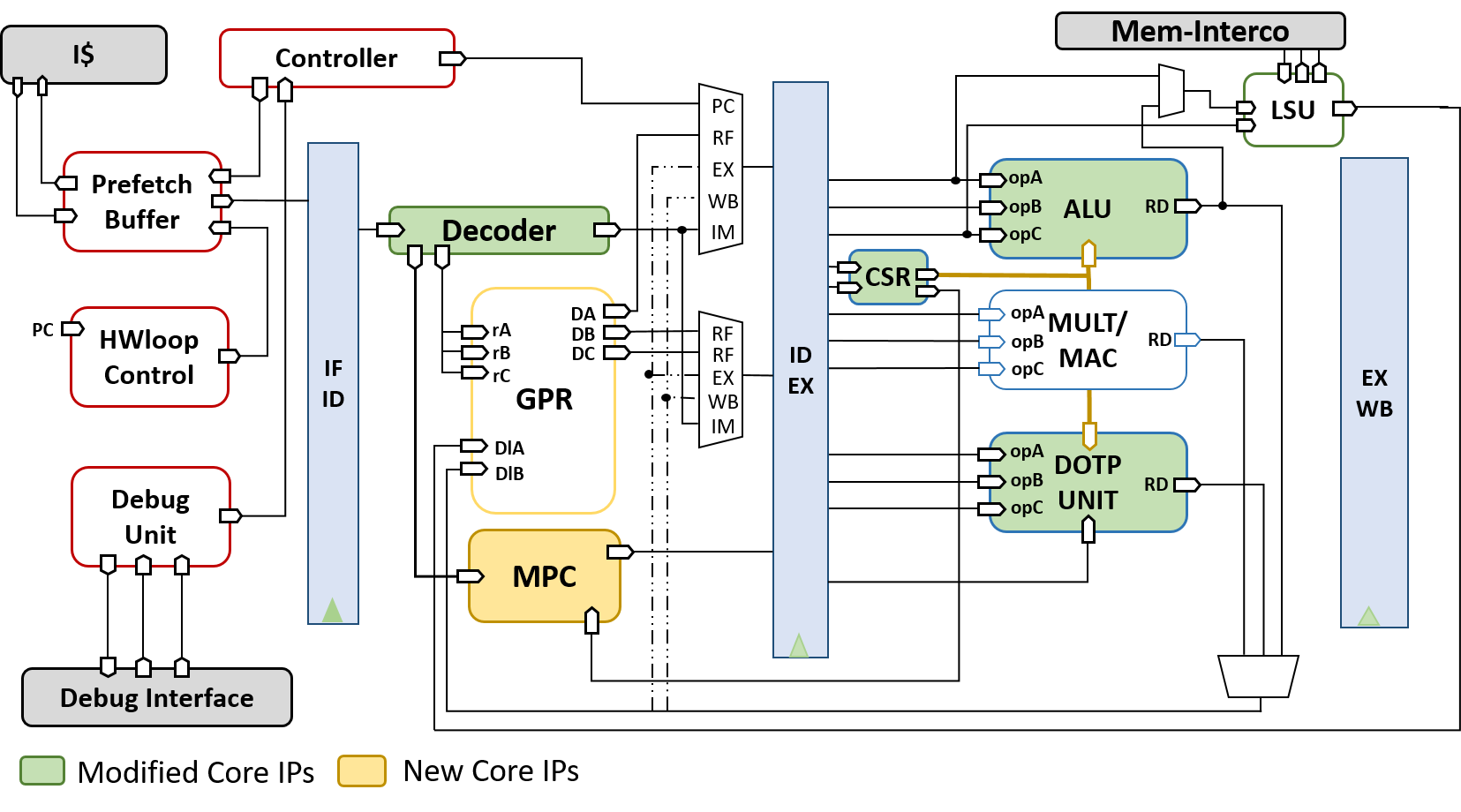}}
\caption{Pipeline of the MPIC core.}
\label{fig:sb_pipeline}
\end{figure}

Figure~\ref{fig:sb_pipeline} shows the diagram of the MPIC pipeline, highlighting the IPs modified to implement the new extension. The logic required to decode the format for SIMD instructions has been removed by the decoder and moved to the CSR, which now has a register dedicated to this task (the orange signal feeds the precision to the functional units). Moreover, two additional registers have been added for managing mixed-precision operations.

\begin{figure}[t]
\centerline{\includegraphics[width=0.5\textwidth]{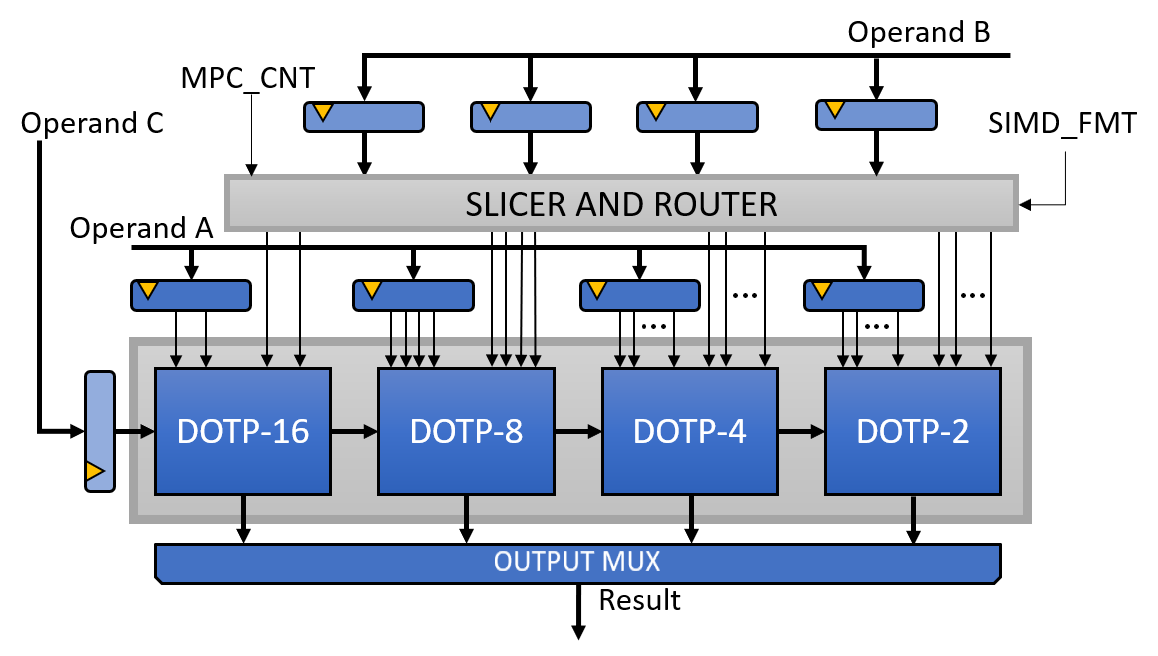}}
\caption{Extended dot-product block.}
\label{fig:dot_product}
\end{figure}

\textit{\textbf{Dot-product Unit:}}
\label{sec:multipliers}
The baseline RI5CY core already supported the cumulative dot product operation for 16- and 8-bit MAC operations; it consists of two distinct sets of multipliers, one for each data size~\cite{RI5CY}. The intermediate multiplications are fed to an adder tree to sum all the contributions. It accepts two 16-bit or four 8-bit operands packed in one 32 bit register, and an optional third input register used as an accumulation register. To extend its support to 4 and 2 bits, we followed the same principle, adding another set of multipliers and an adder tree for each supported format. This configuration enables the execution of 8 and 16 operations per cycle for 4- and 2-bit, respectively, paying the cost for more area but with no impact on the critical path of the design. On the other hand, we apply a power management policy for the unused SIMD units by means of clock gating of the input registers of the units not involved in the current computation, as shown in Figure~\ref{fig:dot_product}.

For what concerns mixed-precision operation, we have operands with different size multiplied together; this implies that one of the input registers contains a higher number of operands than the other, and when we execute a dot-product, one of the input registers is fully utilized, while only a part of the second one is needed. This requires two actions: first, we need to select a sub-group of the second input register (we call it \textit{input B}); second, that sub-group has to be fed to the correct size multiplier; e.g., if we consider an \textit{8x4} MAC, the correct sub-group of operands from input B has to be routed to the 8-bit multiplier. In Figure~\ref{fig:dot_product}, we depict the whole dot-product module. The \textit{slicer and router} block is used to select and direct the correct sub-group of operands to the various \textit{dotp} multipliers; it also sign-extends the smaller operands to match the size of the larger operands. This block is controlled by two signals: \textit{MPC\_CNT} is used to select the sub-groups of operands (discussed later), and \textit{SIMD\_FMT} specifies which type of operands to select (taken directly from the CSR).

A mechanism to correctly choose the sub-group of inputs for register B is needed, so we designed a small controller dedicated to this task.

\textit{\textbf{Mixed-Precision Controller:}}
\label{sec:MPC}
To behave as shown in Fig.~\ref{fig:mixed-matmul}, the mixed-precision controller (MPC) contains a counter that is used to select which subgroup of operands to use. The counter is increased only if the following requisites are met: the \textit{ID\_STAGE} is decoding, which avoids the counter to go up in case of stalls; the instruction is a MAC; the format set by the CSR is mixed-precision. The counter resets by itself depending on the current format used (for example, 8x4 counts up to 2, while 8x2 counts up to 4). However, to implement the execution model explained in Sec.~\ref{sec:execution_model}, a single counter is not enough due to data reuse. This causes each sub-group of operands to be used multiple times before switching to the next one. To work around this problem, we added a second counter that can be programmed to count the number of MACs to execute before changing the sub-group of operands (e.g., in the kernel 4x2 we execute 8 MAC before switching to the next sub-group).
The value of the sub-group is also written inside the CSR; this can be changed by writing directly to it, making it possible to choose the group of operands manually.

\subsection{Compiler support}
We integrated the support to the mixed-precision instruction semantics into the PULP GNU toolchain\footnote{https://github.com/pulp-platform/pulp-riscv-gnu-toolchain}.
The GCC front-end does not require any modification since programmers use a single integral type (i.e., \texttt{int}) and possibly modify the precision of the operations setting the status register; this approach is suitable for both homogeneous and mixed-precision operations.
At the middle-end level, we disabled automatic loop unrolling for the loops that include mixed-precision instructions, intending to avoid inconsistencies with the internal counter of the mixed-precision controller. For the same reason, we inhibited the reordering of the mixed-precision instructions into the compiler backend.

\section{Results}
The evaluation of the MPIC has been performed on two fronts. The first one is the physical implementation, where we extracted values of power, area, and frequency, which were used to compare our approach to the baseline core. The second front was the performance assessment with benchmarks, where we executed a QNN layer from 8 bit down to 2 using uniform and the various mixed-precision variants, also providing a comparison with a commercially available MCUs sporting Cortex M7 and Cortex M4 cores.

\subsection{Implementation Results}
The experimental results presented in this section are based on both RI5CY and MPIC cores integrated into the PULPissimo SoC, which features a full set of peripherals, a DMA subsystem, and 512 kB of SRAM memory. We synthesized the SoC with Synopsys Design Compiler-2016.03 using Global Foundries 22 nm FDX technology, place \& route was performed with Cadence Innovus-15.20.100 in the worst-case corner (SSG 0.59v, -40°C/125°C); power analysis was done at 250 MHz with Typical Corner 0.65V at 25°C using Synopsis PrimeTime. We performed different runs, one to test the max frequency of the SoC, while the other we set the constraint at 250 MHz aiming at maximizing energy efficiency for power analysis purposes.

From the results of the max-frequency synthesis run, we observed a negligible reduction in the maximum operating frequency from 511 to 505 MHz (1\% slowdown). For what concerns power analysis, four different scenarios have been profiled: one to evaluate the impact of the introduced extensions on general-purpose code, while the other 3 while using the modified dot-product unit while executing 8-, 4- and 2-bit QNN kernels.

The results are reported under the power consumption section in Tab. \ref{aap_res}. Surprisingly, MPIC consumes slightly less power in general-purpose applications, thanks to the addition of clock-gating for unused dotp modules (Sec. \ref{sec:multipliers}), which was shared among all the dotp units in RI5CY. Overall, the table shows that power results are all 2\%  within each other, well inside the margin of error, telling us that the changes made did not significantly impact the overall efficiency of the core.

\begin{table}[t]
\begin{center}
\scalebox{1}[1]{\begin{tabular}{|c|c|c|c|}
    \hline
        \multicolumn{4}{|c|}{\textbf{Power consumption results [$mW$] @250 MHz}} \\
    \hline
    & \textbf{RI5CY} &  \textbf{MPIC} & Overhead\\
    \hline
    \textbf{GP App} & 5.39 & 5.30 & -2\%\\
    \hline
    \textbf{8-bit MatMul} & 5.36 & 5.39 & 1\%\\
    \hline
    \textbf{4-bit MatMul} & - & 5.46 & 2\% \\
    \hline
    \textbf{2-bit MatMul} & - & 5.38 & 0\% \\
    \hline
    \hline
    \multicolumn{4}{|c|}{\textbf{Area Results [$\mu m^2$]}} \\
    \hline
    & \textbf{RI5CY} &  \textbf{MPIC} & Overhead\\
    \hline
    \textbf{SoC} & 1002681 & 1004273 & 0.2\%\\ 
    \hline
    \textbf{core} & 15755 & 17584 & 11\%\\  
    \hline
    \textbf{ex\_stage} & 6592 & 7655 & 16\%\\
    \hline
    \textbf{id\_stage} & 5276 & 5673 & 7.5\%\\
    \hline
    \multicolumn{4}{c}{}
\end{tabular}}
\caption{Implementation results in area and power}
\label{aap_res}
\end{center}
\vspace{-5mm}
\end{table}

The second section of Table \ref{aap_res} reports area results. The core has an 11\% overhead given by the extension of the core for status-based operation. The ex\_stage has a 16\% overhead for the added logic to support the new precision operations, while the id\_stage is larger by 7.5\%; this effect is due to several factors: the main contribution is due to the registers that have been added for operand isolation (the id\_stage contains the ID/EX pipeline stage) for MAC operations, while a secondary contribution is that of the mixed-precision controller. Overall, the SoC overhead is around 0.2\% since the 512 kB of SRAM occupies the most area.

\subsection{Benchmarking}
\label{bench}

To show the performance benefits of supporting these new precision formats, we choose a QNN layer with different configurations for input/output and weights. The input tensor size is 16x16x32, while the filter is 64x3x3x32; this configuration is among the ones featuring the best performance on the targeted architectures.
The devices used for this comparison are the baseline core RI5CY, MPIC, STM32H7 equipped with a Cortex M7 (40 nm technology)\cite{stm32h7-datasheet}, and STM32L4 with Cortex M4 (90 nm); results consider the STM32H7 running at 480MHz and STM32L4 at 80MHz \cite{stm32l4-datasheet}.

\begin{figure}[t]
\centerline{\includegraphics[width=0.5\textwidth]{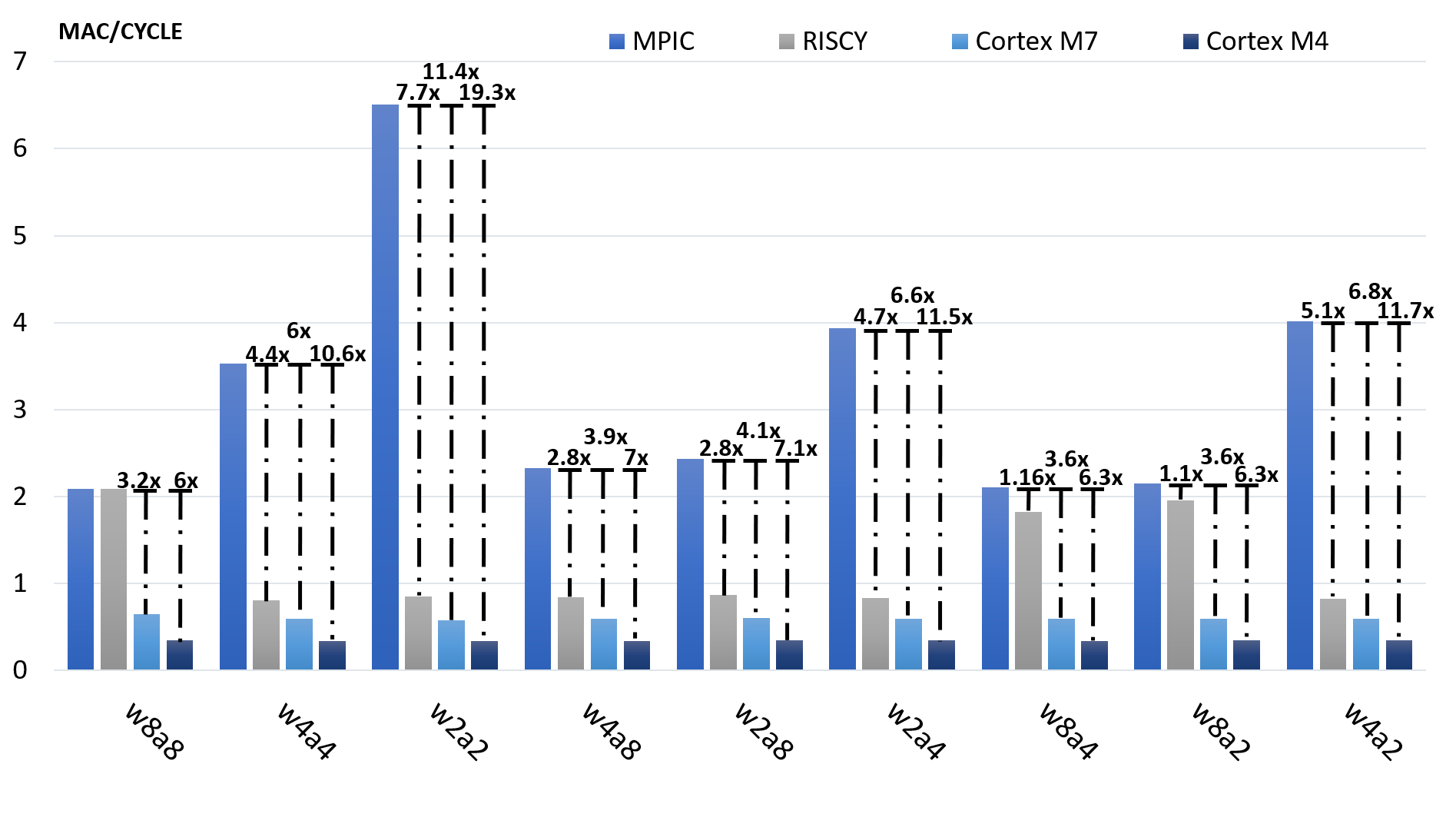}}
\caption{Performance expressed in MAC/Cycle.}
\label{fig:mac_per_cycle}
\end{figure}

Figure~\ref{fig:mac_per_cycle} shows the results in terms of MACs per Cycle.
The different configurations in the chart are denoted in the \textit{X}-axis by the size of the activations and weights. Analyzing the charts, we can see different trends. The Cortex M7 and M4 have lower performance compared to the MPIC in all the configurations, or even in comparison with RI5CY core (in the case of 8-bit uniform quantization or 8-bit weights). This is due to the high overhead introduced by the unpacking of data before the execution of MAC, but also to the fact that only up to two 16-bit MACs in one cycle are supported for both ARM Cortex. 

The RI5CY core presents about 2.1 MAC/Cycle for the first case, well above the Cortex M7/M4 and on par with MPIC, because both support 8-bit MACs. When going to sub-byte configurations, it suffers the same fate as the ARM (except for 8-bit weights) core due to the additional overhead introduced by unpacking data in the \textit{MatMul} phase (Sec. \ref{sec:execution_model}).  We can see that compared to the Cortex M7 or M4, the better performance of RI5CY is due to more efficient ISA: load/store post-increment, hardware loops, and the possibility to execute 4 MACs per cycle at 8-bit precision. In contrast, MPIC does not require to unpack data before execution; data can be fed directly to the dot-product unit, resulting in a peak of 6.5 MACs per cycle in the 2-bit uniform layer. When looking at 8-bit weights, we can see that the performance is close to the 8-bit uniform quantization; this is because unpacking is done in the \textit{im2col} execution phase, which is way less computationally intensive and does not impact execution as much as the inner-loop of the kernel \cite{garofalo2020pulp}.  

Significantly, mixed-precision QNN kernels also do not suffer any performance hit, thanks to unpacking done in hardware. The performance of 8x4 and 8x2 kernels are close to the 8-bit uniform kernel, likewise for the 4-bit one. This is because the selection of the dotp module (Fig. \ref{fig:dot_product}) is tied to the size of the greater operand (e.g., 8x4 uses 8-bit multipliers). However, we can see that the performance is slightly better than their equivalent uniform case, thanks to the higher operational intensity. If we perform a mixed-precision 8x4 operation, operand b needs to fetch fewer data from memory, since its register can hold twice as many operands as the register containing a. 
Another factor that impacts mixed-precision operation is the quantization process (\textit{QntPack}). Focusing on the chart for activations of 4- and 2-bit, the performance is marginally worse than when we have 8-bit activations.

\begin{figure}[t]
\centerline{\includegraphics[width=0.5\textwidth]{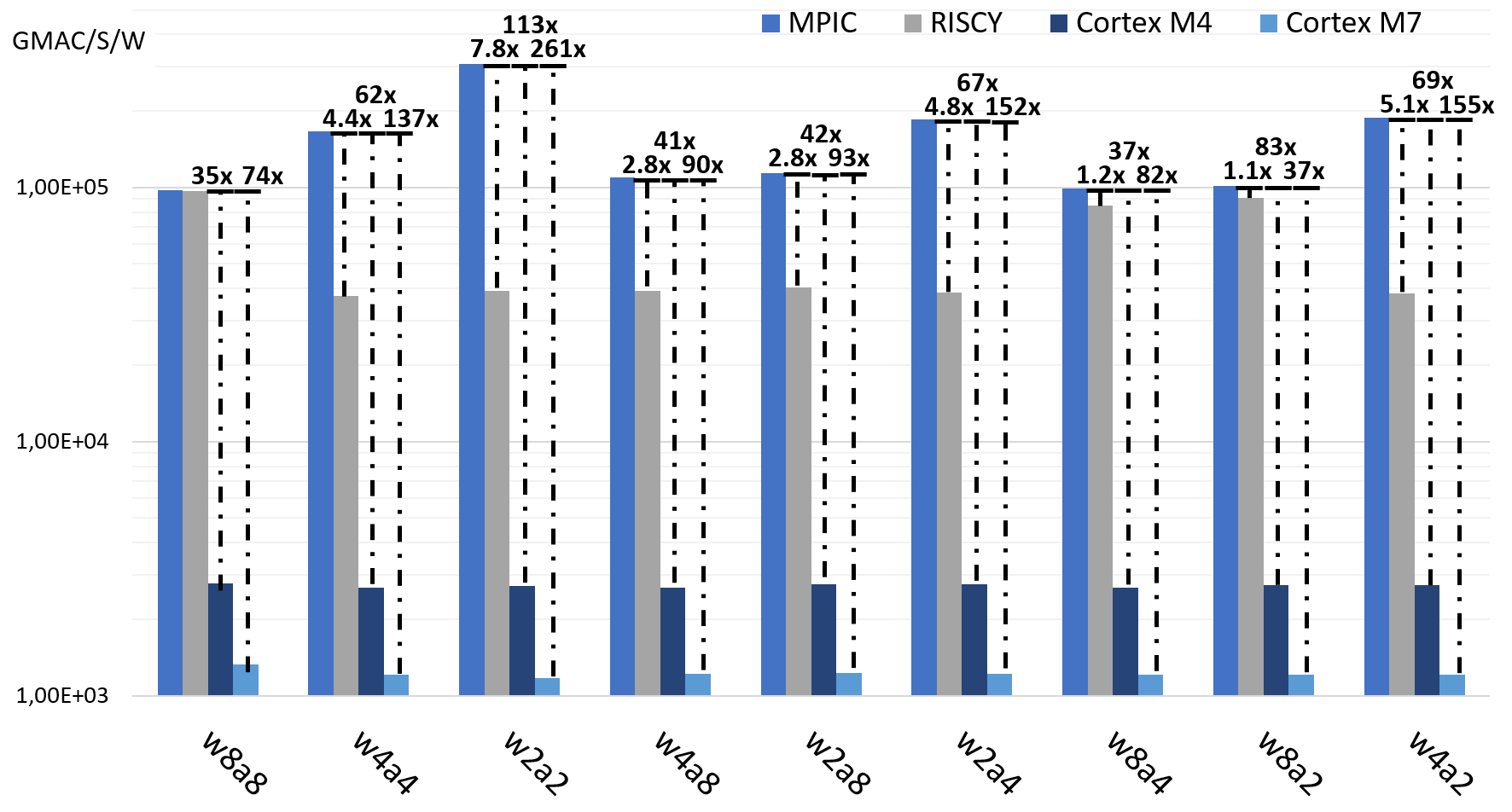}}
\caption{Energy efficiency expressed in GMAC/s/W.}
\label{fig:effi_gmac_per_second_per_watt}
\end{figure}

In contrast with performance in MAC/cycle, energy efficiency (expressed in GMAC/s/W) takes into account also physical design parameters such as the fabrication technology and the operating voltage and frequency.
For the Cortex M7 and M4, we used an implementation from ST-Microelectronics consuming $\sim$234 mW at 480 MHz~\cite{stm32h7-datasheet} and 10 mW at 80 MHz~\cite{stm32l4-datasheet}, respectively; while we used the power consumption figures reported in Table~\ref{aap_res} for the RISC-V SoCs. In Figure~\ref{fig:effi_gmac_per_second_per_watt}, we can see that the lower performance of the Cortex M7 is emphasized even more by the technology factor, having a peak of 1.27 GMAC/s/W and being from 74x to 255x less efficient in these workloads compared to MPIC. The Cortex M4 is way more efficient than the Cortex M7 but still falls short when compared to RISC-V cores, being from 35x to 113x less efficient. For the RI5CY core, we have a slight disadvantage of 1\% only in the 8-bit case, while in all other scenarios, the results are qualitatively similar to the performance ones. 

\section{Conclusion}
In this work, we presented an alternate way to deal with a saturated encoding space. We extended the ISA to support sub-byte and mixed-precision formats aiming at improving the performance of QNN via removing the overhead caused by unpacking data before computation. The MPIC-based SoC implementation resulted in an area overhead of 11\% when compared to the baseline core while having a negligible impact on frequency and power and so not compromising the general-purpose nature of the RI5CY core. The performance gain ranges from 1.1$\times$ to 7.7$\times$ when compared to the baseline during the execution of a QNN layer, and from 3.6$\times$ up to 19.3$\times$ in regard to the Cortex M7 and M4. The energy efficiency peaks at 303 GMAC/s/W for the 2-bit convolution and ranges from one to two orders of magnitude higher when compared with ARM counterpart, providing a solution that is considerably more efficient than commercially available MCUs solutions for QNN inference.

\\
\bibliographystyle{IEEEtran-mod}
\bibliography{bibliogrfia.bib}

\end{document}